# Delay Analysis of Base Station Flow Table in SDN-enabled Radio Access Networks

Seyed Hamed Rastegar, Aliazam Abbasfar, Vahid Shah-Mansouri

*Abstract*—Future generation wireless networks are designed with extremely low delay requirements which makes even small contributed delays important. On the other hand, software defined networking (SDN) has been introduced as a key enabler of future wireless and cellular networks in order to make them more flexible. In SDN, a central controller manages all network equipments by setting the match-action pairs in flow tables of the devices. However, these flow tables have limited capacity and thus are not capable of storing the rules of all the users. In this paper, we consider an SDN-enabled base station (SD-BS) in a cell equipped with a limited capacity flow table. We analyze the expected delay incurred in processing of the incoming packets to the SD-BS and present a mathematical expression for it in terms of density of the users and cell area.

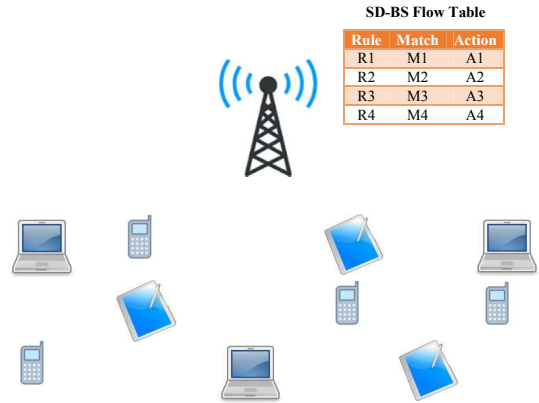

Figure 1. An SD-BS equiped with a TCAM-based flow table serving cellular users

## I. Research Motivation

SDN-enabled devices employ Ternary Content Addressable Memories (TCAMs) in their flow tables for high speed packet processing. However, TCAMs are very expensive and consume much power [1]. Therefore, SDN-enabled devices are equipped with limited TCAM resources which is not capable of storing the rules of all users. This results in a delay for processing the incoming packets to these devices. Analysis of this delay have been so far a topic of research in wired networks [2], [3]. Nonetheless, this problem has not been well studied for the case of wireless and cellular networks. To fill this gap, in this paper we consider an SDN-enabled base station (SD-BS) in a cell equipped with a limited capacity flow table. We analyze the expected delay incurred in processing of the incoming packets to the SD-BS and present a mathematical expression for it in terms of density of the users and cell area.

## II. System Scenario

In this section, we describe our considered system model followed by problem formulation.

### A. System Model

We consider radio access network (RAN) part of an SDN-enabled cellular network [4], [5] where different equipments operate based on the received commands from the centralized SDN controller. According to Figure 1, we focus our attention to a single cell where $N$ users are associated with a software defined BS (SD-BS) to which they are sending their data flows.

The SD-BS is equipped with a flow table contains match-action pairs for different flows as in Figure 1. The entries in the table are resulted from the commands which previously has been sent by the SDN controller to the SD-BS and stored in the flow-table. Due to high speed processing requirements, we assume the flow table is a TCAM-based one [6] with limited capacity only capable of storing $C$ match-action pairs (rules) at the same time.

In our considered scenario, when a packet of a user's flow is arrived at the SD-BS three possible choices might occur. If the header is consistent with one of the match fields in the flow table and thus the related rule is already available, the SD-BS will perform required action indicated. Otherwise, the SD-BS will send the packet information to the SDN controller which in response sends back required actions to be applied. In this case, if the flow table is not full, the action will be cached in the table in order to be used for subsequent packets of this flow. However, if the table is full, the entry will not be cached and therefore it is needed to ask again the controller for the next packets of this flow, which incurs some delay [7]. We consider a time slotted scenario where at each time slot the SD-BS randomly (with equal probability) selects $C$ out of $N$ users in the cell and store their rules in the flow table, thus SD-BS should fetch the rules of packets belonging to remaining $N - C$ users from the controller.

## B. Problem Formulation

As stated in the system model, considering a TCAM-based flow table capable of storing $C$ rules simultaneously where in each time slot the entries of flow table are randomly allocated with equal probabilities to $N$ users, the average delay for processing a packet in the SD-BS will be

$$\text{Delay} = \begin{cases} \left(1 - \frac{C}{N}\right) \times d_{\text{ctrl}} & \text{if } N > C \\ 0 & \text{if } N \leq C \end{cases} \quad (1)$$

where $d_{\text{ctrl}}$ is the time required to fetch a packet's rule from the controller. In (1), if $N \leq C$ there are enough spaces to store the rules of all users while for $N > C$ the rule of a packet is not available in the flow table with probability $1 - C/N$ which causes delay of $d_{\text{ctrl}}$. Also, we have not considered the delay for rule look up in TCAM-based flow table since it is negligible in comparison to $d_{\text{ctrl}}$. However, the parameter $N$ in (2) is a random variable and we should use the expected value of Delay as

$$\text{E[Delay]} = \sum_{n=C+1}^{\infty} \left[\left(1 - \frac{C}{n}\right) \times d_{\text{ctrl}} \times \Pr\{N = n\}\right]. \quad (2)$$

In the next section, we will mathematically derive the above expectation.

## III. EXPECTED DELAY CALCULATION

To find the expectation in (2), first we should know the probability mass function (pmf) of the number of users in a cell, and then determine the required summation which we investigate in the following.

As stated in [8], assuming users locating under widespread Poisson Point Process (PPP) model with density $\lambda_u$, for both homogeneous and heterogeneous cellular network models, the number of users in a cell with area $A$ is distributed according to the following pmf

$$P_A(N = n) = \frac{(\lambda_u A)^n}{n!} \exp(-\lambda_u A), \quad (3)$$

which is a Poisson distribution with parameter $\lambda_u A$.

Using the pmf in (3), the expected delay of (2) can be written as

$$\text{E[Delay]} = \sum_{n=C+1}^{\infty} \left[P_A(N = n) - \frac{C}{n} \times P_A(N = n)\right] d_{\text{ctrl}} =$$

$$\left[(1 - \Pr\{N \leq C\}) - C \sum_{k=C+1}^{\infty} \frac{1}{k} \frac{(\lambda_u A)^k}{k!} \exp(-\lambda_u A)\right] d_{\text{ctrl}}, \quad (4)$$

in which $\Pr\{N \leq C\}$ is the cumulative distribution function (CDF) of random variable $N$ in (3) and is equal to $\frac{\Gamma(C+1, \lambda_u A)}{C!}$ where $\Gamma(m, z)$ is the incomplete Gamma function defined as [9]

$$\Gamma(m, z) \triangleq \int_z^{\infty} t^{m-1} e^{-t} dt. \quad (5)$$

Therefore, the remaining challenge to calculate the expected delay is to determine the summation in (4) which we restate here as a function of new variable $b \triangleq \lambda_u A$:

$$E_C(b) = \sum_{k=C+1}^{\infty} \frac{1}{k} \frac{(b)^k}{k!} \exp(-b). \quad (6)$$

In order to determine the above sum, we take the derivative of both sides of (6) as

$$\frac{dE_C(b)}{db} = \sum_{k=C+1}^{\infty} \frac{1}{k} \frac{kb^{k-1}}{k!} \exp(-b) - \sum_{k=C+1}^{\infty} \frac{1}{k} \frac{b^k}{k!} \exp(-b)$$

$$= \sum_{k=C+1}^{\infty} \frac{b^{k-1}}{k!} \exp(-b) - E_C(b). \quad (7)$$

However due to the fact that for the pmf in (3), $\sum_{n=0}^{\infty} \frac{(b)^n}{n!} \exp(-b) = 1$, the sum in (7) is written as

$$\sum_{k=C+1}^{\infty} \frac{b^{k-1}}{k!} \exp(-b) = \frac{1}{b} \left(1 - \sum_{k=0}^{C} \frac{b^k}{k!} \exp(-b)\right). \quad (8)$$

Taking together (7) and (8), we will arrive at the following differential equation:

$$\frac{dE_C(b)}{db} + E_C(b) = \frac{1}{b} \left(1 - \sum_{k=0}^{C} \frac{b^k}{k!} \exp(-b)\right)$$

$$E_C(b = 0) = 0, \quad (9)$$

in which the boundary condition has been obtained by setting $b = 0$ in (6). In order to solve this differential equation, we multiply both sides by $\exp(b)$ as

$$\exp(b) \frac{dE_C(b)}{db} + \exp(b) E_C(b) = \frac{1}{b} \left(\exp(b) - \sum_{k=0}^{C} \frac{b^k}{k!}\right), \quad (10)$$

however the left side of the above equation can be written as $\frac{d(E_C(b) \exp(b))}{db}$. Now taking the integral of both sides of (10), the solution of (9) is obtained as

$$E_C(b) = \left[\int_1^b e^t t^{-1} dt - \ln(b) - \sum_{k=1}^{C} \frac{b^k}{k \times k!} + \beta\right] e^{-b}. \quad (11)$$

where $\beta$ is a constant. To calculate $\beta$, we should use the boundary condition $b(0) = 0$, however, since the term $\int_1^b e^t t^{-1} dt - \ln(b)$ is indeterminate at $b = 0$, we use the L'Hospital's rule around $b = 0$ which results in $\beta = -1$. To write (11) in a more suitable form, we exploit the exponential integral function defined as [9]

$$\text{Ei}(z) \triangleq \int_{-\infty}^{z} e^t t^{-1} dt, \quad (12)$$

which is a well known function available in most numerical software packages and we can write the integral in (11) as $\int_1^b e^t t^{-1} dt = \text{Ei}(b) - \text{Ei}(1)$.

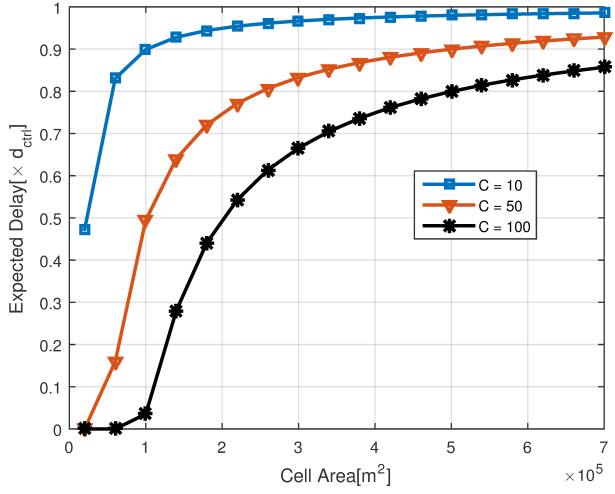

Figure 2. Plot of Expected Packet Delay versus Cell Area ($A$) for $\lambda_u = 10^{-3}/m^2$

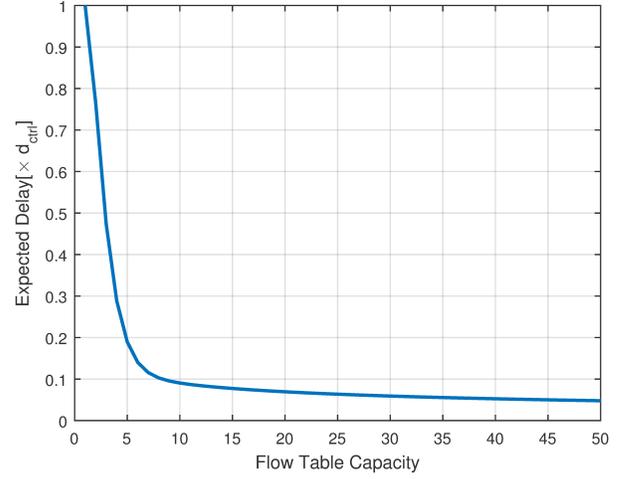

Figure 3. Plot of Expected Packet Delay versus Flow Table Capacity for $C = \mathrm{E}[N]$

Putting all above findings together, we can now state our final result: Assuming a cell with area $A$ and users distributed according to a PPP with density $\lambda_u$, using a TCAM-based flow table with limited capacity of $C$ causes transmitted packets to experience the following expected delay:

$$\mathrm{E}[\mathrm{Delay}] = \left(1 - \frac{\Gamma(C+1, \lambda_u A)}{C!} - C\left\{\mathrm{Ei}(\lambda_u A) - \mathrm{Ei}(1) - \ln(\lambda_u A) - \sum_{k=1}^{C} \frac{(\lambda_u A)^k}{k \times k!} - 1\right\} e^{-\lambda_u A}\right) d_{\mathrm{ctrl}}. \quad (13)$$

## IV. NUMERICAL RESULTS

In this section, we present some numerical results for our derived expression of expected delay. All the required functions in (13) are built-in in most numerical softwares such as MATLAB thus we can efficiently determine the value of expected delay. Also, we set no value for delay from SD-BS to the controller ($d_{\mathrm{ctrl}}$), since the expression in (13) is a multiplication of this parameter and we plot the expected delay normalized to $d_{\mathrm{ctrl}}$. First, we have considered a cell in a cellular network where users are distributed according to a PPP model with density $\lambda_u = 10^{-3}/m^2$. We plotted expected delay versus the cell area ($A$) considering three different capacities of $C = 10, 50, 100$ for the base station flow table as in Figure 2. According to this figure, the expected delay increases by increasing cell area and decreasing flow table capacity. In the second plot, we have proceeded an other approach. We set parameter $C = \mathrm{E}[N] = \lambda_u A$ in (13) and plotted the expected delay versus $C$ in Figure 3. This is an intuitive choice to use flow table capacity equal to the expected number of users in the cell. In this plot, delay decreases by increasing $C$, however this decrease is much more significant for $C \in [1, 10]$ and then becomes slow. It is worth nothing that by setting $C = \lambda_u A$, the expression for expected delay in (13) will be just a function of $C$ and due to linear relation of $\lambda_u$ and $A$ with $C$, plotting the expected delay versus either of these two parameters will have the same results as Figure 3.

## V. CONCLUSION

In this paper, we consider an SDN-enabled radio access network where the base stations are equipped with limited TCAM-based flow tables. Using tools from differential equations theory, we determine closed form expression for expected delay of transmitted packets by the users in a cell. The expression gives a good insight to network designers about the relation between flow table capacity and delay which can be used to find how much TCAM is needed in base stations in order to have a bounded delay.